# Overview of the high-definition x-ray imager instrument on the Lynx x-ray surveyor

Abraham D. Falcone,[a,*] Ralph P. Kraft,[b] Marshall W. Bautz,[c] Jessica A. Gaskin,[d] John A. Mulqueen,[d] and Doug A. Swartz,[d] for the Lynx Science and Technology Definition Team
[a]Pennsylvania State University, Department of Astronomy and Astrophysics, University Park, Pennsylvania, United States
[b]Harvard-Smithsonian Center for Astrophysics, Cambridge, Massachusetts, United States
[c]Massachusetts Institute of Technology, Kavli Institute for Astrophysics and Space Research, Cambridge, Massachusetts, United States
[d]NASA Marshall Space Flight Center, Huntsville, Alabama, United States

**Abstract.** Four NASA Science and Technology Definition Teams have been convened in order to develop and study four mission concepts to be evaluated by the upcoming 2020 Decadal Survey. The Lynx x-ray surveyor mission is one of these four large missions. Lynx will couple fine angular resolution (<0.5 arcsec HPD) x-ray optics with large effective area (∼2 m$^2$ at 1 keV), thus enabling exploration within a unique scientific parameter space. One of the primary soft x-ray imaging instruments being baselined for this mission concept is the high-definition x-ray imager, HDXI. This instrument would use a finely pixelated silicon sensor array to achieve fine angular resolution imaging over a wide field of view (∼22 × 22 arcmin). Silicon sensors enable large-format/small-pixel devices, radiation tolerant designs, and high quantum efficiency across the entire soft x-ray band-pass. To fully exploit the large collecting area of Lynx (∼30× Chandra), with negligible or minimal x-ray event pile-up, the HDXI will be capable of much faster frame rates than current x-ray imagers. We summarize the planned requirements, capabilities, and development status of the HDXI instrument, and associated papers in this special edition will provide further details on some specific detector options. © *The Authors. Published by SPIE under a Creative Commons Attribution 4.0 Unported License. Distribution or reproduction of this work in whole or in part requires full attribution of the original publication, including its DOI.* [DOI: 10.1117/1.JATIS.5.2.021019]

## 1 Introduction

The Lynx x-ray surveyor is one of four large missions currently being studied by a NASA Science and Technology Definition Team (STDT). At the conclusion of this study, the team will have developed a concept, along with technology development plans and cost estimates, as input to the 2020 Decadal Survey process. Lynx is a concept for an x-ray observatory that will directly observe the dawn of supermassive black holes, reveal the invisible drivers of galaxy and structure formation, and trace the energetic side of stellar evolution and stellar ecosystems. It will provide huge improvements in x-ray sensitivity relative to existing and planned x-ray missions, including Athena, Chandra, and XMM. The current notional design for Lynx utilizes nested x-ray mirrors providing ∼2.3 m$^2$ effective area at 1 keV. The angular resolution of the mirror system is expected to be better than 0.5 arcsec on-axis and better than 1 arcsec within a 10-arcmin radius at the focal plane. The combination of this large effective area and fine angular resolution leads to an ∼100× improvement in sensitivity relative to existing and planned missions. The fine angular resolution across the field of view enables deep surveys with little to no source confusion, enabling, e.g., the detection of very distant and dim AGN out to redshifts of $z \sim 10$. These unprecedented characteristics, along with the excellent spectral resolution for point-like and extended sources, will enable: detailed studies of the first seed black holes, as well as their growth and evolution with time in the universe, the impact of black hole system emission/feedback on multiple scales, the mapping of the intergalactic cosmic structures and the study of its impact on galactic growth and evolution, detailed studies of galaxy clusters, studies of stellar evolution and birth, along with the impacts of feedback on stellar systems, and the impact of x-ray stellar activity on extrasolar planet habitability. In addition to the above science topics, Lynx will be a large observatory class mission, which would enable myriad major-impact community-driven science studies through a vibrant general observer program. For a more thorough overview of the Lynx mission and its science motivation, refer to Ref. 1.

The current baseline Lynx design includes three instruments: (1) the high-definition x-ray imager (HDXI), (2) the Lynx x-ray microcalorimeter (LXM), and (3) the x-ray grating spectrometer (XGS). The HDXI and the LXM are mounted on a translation table (see Fig. 1), which allows one of them to be in the field of view at any given time. The grating array readout is mounted to the side and is always in position for readout of the diffracted and offset x-rays, with its utility dependent on whether or not the grating itself is in position. The LXM is discussed by Bandler et al.;[2] the XGS is discussed by McEntaffer et al.[3] and Moritz et al.;[4] and in this paper, we will briefly describe the HDXI instrument. The majority of the description in this paper was presented at the SPIE Astronomical Telescopes meeting in 2018 and was reported in the associated proceedings.[5]

## 2 HDXI Overview

The HDXI instrument is an imaging x-ray spectrometer that will achieve a moderately wide field of view while simultaneously

*Address all correspondence to Abraham D. Falcone, E-mail: afalcone@astro.psu.edu

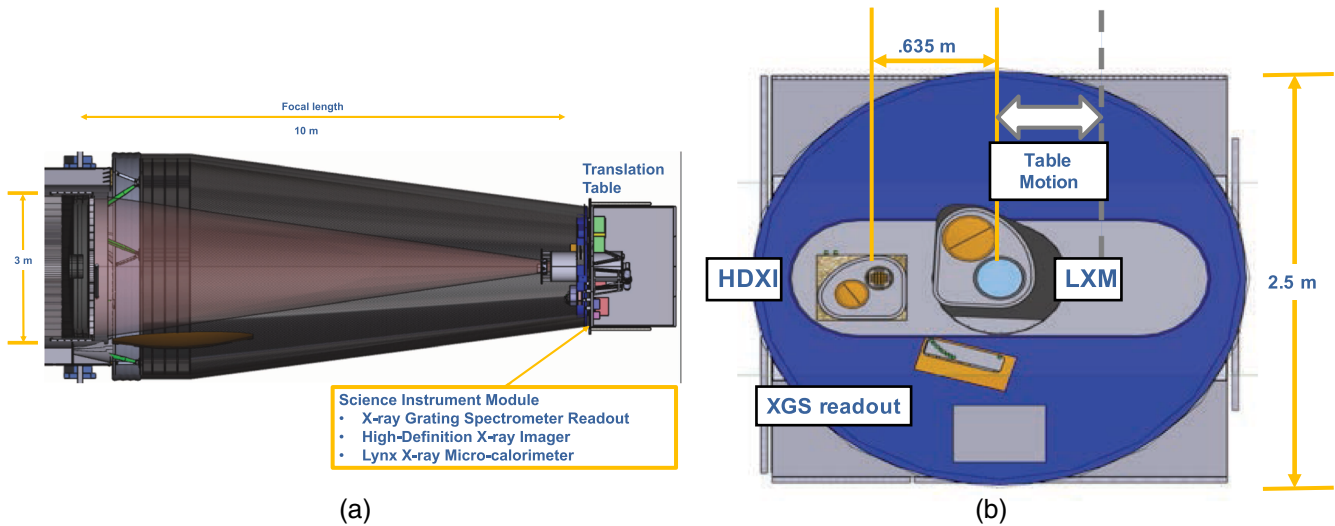

**Fig. 1** (a) Initial/notional design of Lynx x-ray surveyor telescope assembly, image credit to NASA, M. Baysinger; and (b) focal plane layout. The translation table moves HDXI or LXM into position along with the associated support structure, heat pipes, and radiators.

achieving fine angular resolution. In order to adequately oversample the <0.5 arcsec point-spread-function (PSF) of the Lynx mirrors, the HDXI is designed with pixels that span less than 0.33 arcsec per pixel, which corresponds to pixels with a pitch that is ≤16 μm for the baseline focal length of 10 m. This pixel size ensures that individual photon landing locations will be constrained to a region size that is smaller than the PSF, and this region can be constrained to an even smaller size through the use of subpixel positioning when the charge distribution is estimated from the x-ray event grade. With a field of field-of-view requirement driven by the need to efficiently survey the x-ray sky for distant accreting active galactic nuclei, the notional focal plane instrument is designed to be 22 × 22 arcmin. The combination of these two requirements leads to a focal plane camera with ∼16 megapixels that spans ∼6.5 cm × 6.5 cm. The large format, combined with a need for good 0.2 to 10 keV quantum efficiency, moderate spectral resolution, and rapid readout of the high x-ray throughput, leads to silicon detectors as a natural choice for the HDXI focal plane instrument. Since the survey science also requires <1 arcsec resolution near the edges of the field-of-view (FOV), we are baselining multiple detectors to cover this FOV since this allows the detectors to be tilted in order to better match the curvature of the focal surface. The notional design includes 21 silicon detectors, with a layout as shown in Fig. 2. It is also possible that a future concept for this array could include larger detectors that are intrinsically curved but that has not been baselined at this time. One could also consider a future concept that utilizes larger pixels that are capable of subpixel spatial resolution for the x-ray event localization,[6,7] but for simplicity and consistency with past implementation (e.g., Chandra) during this concept study, we have baselined a concept that utilizes pixels that directly oversample the PSF of the optics.

The above-stated HDXI parameters and more detailed instrument requirements are driven by the science requirements derived by the Lynx STDT. These initial requirements and parameters are summarized in Table 1. In addition to the FOV and pixel size requirements that were described above, the high throughput of the x-ray mirrors necessitates fast readout of the detectors in order to avoid pile-up of the individual x-rays that would otherwise lead to loss of energy resolution and flux measurement accuracy. The energy range is driven by the typical soft x-ray flux of the sources required to achieve the prime science goals of the mission.

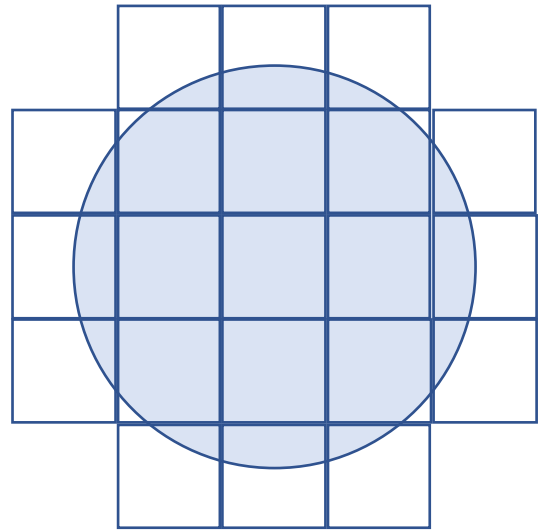

**Fig. 2** Schematic of detector layout. The circle shows the 22-arcmin diameter requirement on the field of view, and the layout of the detectors is shown to cover the field with considerable margin.

The Marshall Space Flight Center Advanced Concept Office (MSFC–ACO) and the Goddard Space Flight Center Instrument Design Lab (GSFC–IDL) have carried out engineering and design studies. These studies have led to significant refinement of the notional design for HDXI, as well as initial estimates of mass, power, volume, and cost. A notional instrument configuration that resulted from these studies is shown in Figs. 3 and 4. This configuration includes the 21-sensor array that is mounted to a mosaic plate, which is cooled to −90°C through a cold strap connection to the instrument chamber housing. This housing, along with its thermal load from the rest of the instrument, is passively cooled via external radiators that are connected to the instrument through heat pipes that move along with the translation table. Thermal modeling indicated that this design was feasible for cooling the detectors, as well as an array of ASICs that were mounted on the front-end mother board.

Table 1 Requirements and baseline parameters table for HDXI.

| HDXI parameter | Requirement | Science drivers |
| --- | --- | --- |
| Energy range | 0.2 to 10 keV | Sensitivity to soft x-ray sources, particularly at high-$z$ |
| Quantum efficiency (excluding filter) | ≥0.85 (0.5 to 7 keV); ≥TBD (0.2 to 0.5 keV) | |
| Field of view | ≥22 × 22 arcmin | Deep survey efficiency $R_{200}$ for nearby galaxies |
| Pixel size | ≤ 16 × 16 $\mu$m | Point source sensitivity; resolve AGN from group emission (actual requirement could evolve to measurement accuracy of x-ray event localization) |
| Read noise | ≤4 e$^-$ | Low-energy detection efficiency |
| Energy resolution (FWHM) | ≤70 eV @ 0.3 keV; ≤150 eV @ 5.9 keV | Low-energy detection efficiency; source and background spectra |
| Full-field count-rate capability | ≥8000 ct/s | No dead time for bright diffuse sources (e.g., Perseus Cluster or case A) |
| Frame rate full-field window (20" × 20") | >100 frame/s; >10,000 window/s | Maximize low-energy throughput, minimize background, minimize pile-up |
| Radiation tolerance | 5 year @ L2 (baseline); 20 year @ L2 (goal) | Maintain science capabilities for >5 to 20 years in orbit |

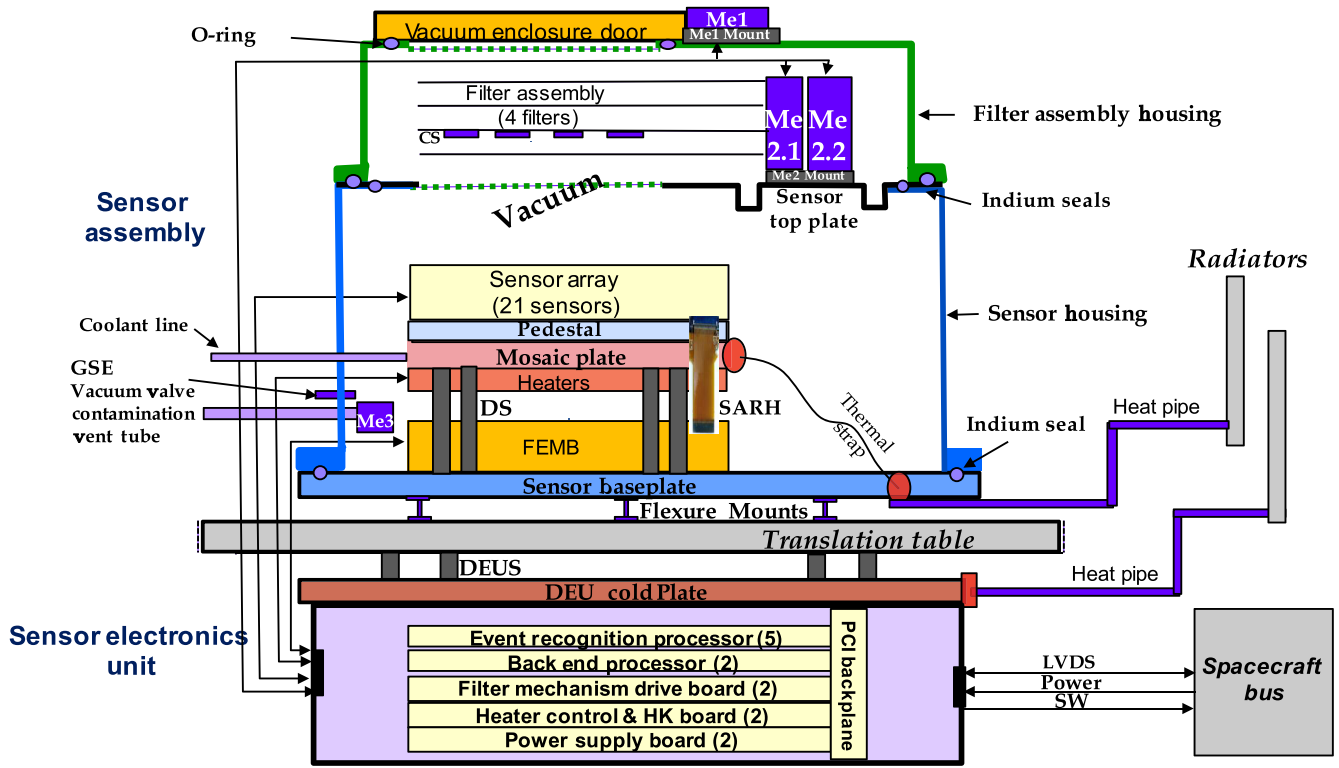

**Fig. 3** Schematic block diagram of HDXI from MSFC-ACO and GSFC-IDL combined studies.

This sensor assembly vacuum chamber also contains an array of retractable filters: two for blocking UV/optical light, one protective blocking filter, and one with calibration x-ray sources. The electronics box contains boards for main processing, power, heater and filter control, housekeeping, and FPGA boards for grading and packaging x-ray event data. These FPGA boards are for event recognition processing, and they operate on parallel data streams to extract the x-ray event pixel data from the full-frames; thus enabling the telemetering of x-ray event data without the need to telemeter full frames in normal observing mode. Occasional full frame data can also be telemetered in order to enable calibration data. Based on these conceptual design studies, we expect the HDXI instrument to have a total power of ∼178 W, a mass of ∼80 kg, and an average telemetry rate of ∼600 kBytes/s. This initial design is intended as a reference point to estimate required resources and feasibility.

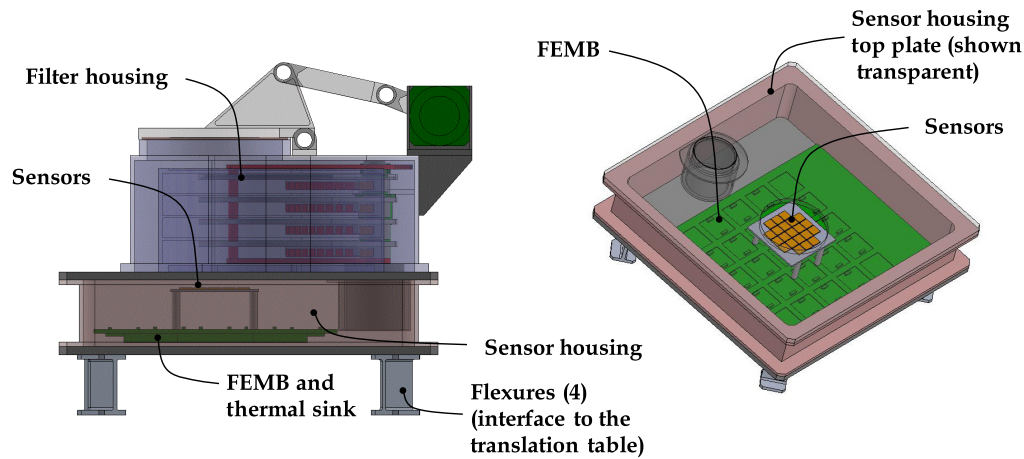

**Fig. 4** Detector assembly diagram for notional HDXI instrument, showing the filter assembly housing (designed at the GSFC - IDL) above the detector plane, which resides inside a vacuum enclosure (shown with transparent walls).

## 3 HDXI Detectors

Three primary detector types are being considered for the notional HDXI instrument design. These detectors include two forms of active pixel silicon sensors, namely x-ray hybrid CMOS detectors (HCDs) and x-ray monolithic CMOS detectors, as well as one form of charge-coupled device (CCD) known as a digital CCD (see Fig. 5). Each of the three sensor technologies has a clear path toward reaching TRL 5 by the time of mission adoption and phase B, which is projected to be in ∼2026, but each of the technologies needs further research and development in order to reach this point.

### 3.1 X-Ray Hybrid CMOS Detectors

X-ray HCDs are active pixel sensors that are made by bonding a silicon detection layer, with an optimized thickness and resistivity, to another silicon layer containing readout-integrated circuitry in each pixel. They are back-illuminated devices offering high quantum efficiency over the full soft x-ray band pass (0.2 to 10 keV) with fully depleted silicon and deep depletion depths (>100 μm). These x-ray detectors are being developed in a collaborative program between Teledyne Imaging Systems and the Pennsylvania State University. These devices are inherently radiation hard due to the fact that the charge is not transferred across the device and, is instead, readout directly from each pixel. They offer very low power relative to CCDs, and they have displayed high readout rates with multiplexed readout lines (>5 MHz per line through 32 lines). X-ray HCDs that were developed for other purposes have been demonstrated to TRL 9 in a recent rocket flight[8] and non-x-ray versions have flown on recent observatories (e.g., OCO-2), but these devices have moderate noise levels and large pixels. Newer test devices have been fabricated and tested with lower noise (∼5.5 e⁻ RMS), small pixels (12.5 μm pitch), and on-chip correlated double sampling.[9] Pixel-to-pixel calibration has also been successfully demonstrated and the test devices have shown energy resolution of 78 eV (FWHM) at 0.5 keV and ∼150 eV (FWHM) at 5.9 keV.[10] The devices have also been operated at substrate bias voltages ranging all the way up to 150 V, which enables the optimization of charge splitting between pixels in order to prioritize less charge splitting for better energy resolution or some slight degree of increased charge splitting in order to obtain improved spatial resolution of the x-ray photon landing location. Some of the recently developed devices are also capable of event driven readout and digitization on-chip. Future developments aim to achieve low readnoise (<4 e−) while retaining the already-achieved small pixels with on-chip correlated double sampling and digitization, deep depletion, high speed, low power, and radiation hardness.

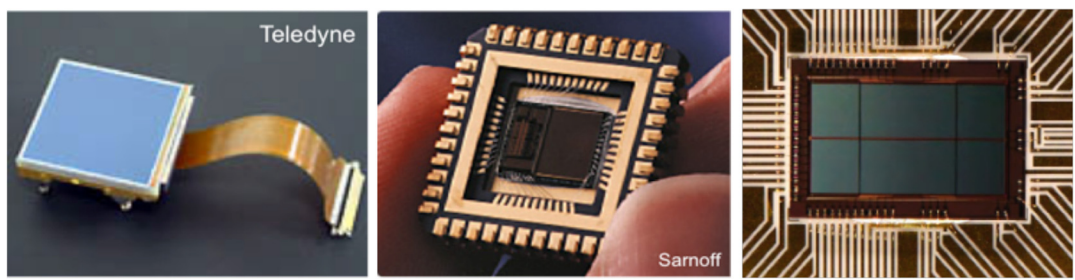

**Fig. 5** Three silicon detector technologies are currently being used to help specify HDXI parameters. They are: (1) hybrid CMOS active pixel sensors being developed by Teledyne and the Pennsylvania State University, (2) monolithic CMOS active pixel sensors being developed by Sarnoff and Harvard-Smithsonian Astrophysical Observatory, and (3) digital CCDs being developed by Lincoln Labs and the Massachusetts Institute of Technology.

### 3.2 Monolithic CMOS Detector

Monolithic CMOS sensors are under development by Smithsonian Astrophysical Observatory and Sarnoff Research Institute for a wide variety of x-ray astrophysics applications, including Lynx. Low noise (∼2.9 e− rms), low dark current, and good sensitivity to low energy x-rays (B $K_\alpha$ and C $K_\alpha$) were demonstrated on back-illuminated 1 k × 1 k 16 $\mu$m pixel pitch sensors built on 10-$\mu$m thick epitaxial silicon.[11] Intrinsic advantages of this technology include radiation hardness, high frame rates, low power, and high levels of on-chip integration. The next generation of sensors under development will be built with PMOS technology with n-type Si substrates.[12] As described in Refs. [12] and [13], the collection of holes instead of electrons has the potential to enable improvements in read noise, dark current, and radiation hardness. PMOS technology potentially offers several other advantages for x-ray astronomy over the conventional NMOS technology. The effect of trapping is greatly reduced at the sensor surface, thus potentially improving the energy resolution below 1 keV for back-illuminated devices. This technology could also virtually eliminate random telegraph signal effects. Preliminary measurements of PMOS sensors achieved ∼2 e− (RMS). X-ray illumination of front-illuminated PMOS sensors shows good energy resolution for $^{55}$Fe x-ray, but they show significantly more charge diffusion (i.e., split pixel events) than NMOS sensors.[12] Future developments aim to achieve deep depletion with the addition of a bias layer to the back surface to increase the quantum efficiency to 10 keV.

### 3.3 Digital CCDs

The digital CCD architecture under development at Massachusetts Institute of Technology (MIT) Lincoln Laboratory aims to realize a large-format, low-noise imager with the sensitivity and uniformity of a scientific-grade CCD and the signal processing and digitization capabilities of CMOS imager. To this end, two parallel development paths are being pursued. The first aims to develop an analog CCD sensor with multiple, high-speed (up to 5 MHz), low-noise (potentially subelectron) output amplifiers capable of efficient charge transfer with voltages compatible with CMOS logic levels. The second path aims to develop compact, highly parallel, low-power signal processing chains allowing high-frame rates with power consumption consistent with the constraints of a space-borne instrument. When fully demonstrated, these two technologies will be combined, first as discrete elements, and then potentially as a single detector module by means, for example, of three-dimensional integration of analog and digital tiers through hybrid wafer bonding. In principal, either of these configurations could be deployed in a flight instrument for Lynx. First-generation digital CCDs achieve noise of 4.6 electrons (RMS) at 2.5 MHz pixel rates and good charge transfer efficiency with CMOS compatible charge transfer clock swings as low as 1 V peak-to-peak.[14,15] NASA has funded a strategic astrophysics technology program to further develop this technology over the coming year. This development effort aims to demonstrate lower-noise, radiation-tolerant versions of these devices.

## 4 Summary

The HDXI instrument on Lynx would enable breakthrough science by providing massive improvements in x-ray source sensitivity through the use of fine angular resolution detectors over a wide enough field of view to enable survey science at the high redshifts and for low luminosity targets. Each of the detector technologies described above is capable of achieving some of the requirements of HDXI, but none of these detector technologies can currently achieve all of the requirements simultaneously. Technology development over the next ∼6 years will bring at least one of these technologies to TRL 5 for the start of the Lynx mission. While other detectors may be capable of achieving the requirements in the future and while further trade studies will be carried out to choose the optimum detector, these three options have enabled the notional instrument to be specified in a way that has proven feasibility. The notional HDXI design has shown that it can accommodate any of these three detector types and that reasonable mass, power, and cost are achievable. Through multiple MSFC-ACO and GSFC-IDL runs, we have developed a notional design for the HDXI instrument, and we have shown that it can be accommodated within reasonable constraints while fulfilling requirements that satisfy the groundbreaking science drivers identified by the Lynx Science and Technology Definition Team.

### Acknowledgments

The authors gratefully acknowledge the efforts of the many scientists that have contributed to this instrument concept as part of the Lynx Science & Technology Definition Team process. The authors are also grateful for the efforts of the Goddard Space Flight Center Instrument Design Lab and the Marshall Space Flight Center Advanced Concepts Office for their contributions to this initial instrument design concept. Much of the material in this manuscript was also presented as an SPIE proceedings.[5]

**Abraham D. Falcone** is a research professor of Astronomy and Astrophysics at the Pennsylvania State University. He received his BS degree in physics (Virginia Tech) in 1995, and his PhD in physics (UNH) in 2001. He has authored more than 200 refereed journal publications. He leads research in x-ray and gamma-ray instrumentation, and studies of high-energy astrophysical sites such as active galactic nuclei, gamma ray bursts, and x-ray binaries. He cochairs the HDXI Instrument working group for Lynx.

Biographies of the other authors are not available.